\documentclass[11pt]{article}

\usepackage{amssymb}
\usepackage{amsmath}
\usepackage{amsfonts}

\newcommand{\R}{\mathbb{R}}

\newcommand{\Z}{\mathbb{Z}}

\newcommand{\be}{\begin{equation}}
\newcommand{\ee}{\end{equation}}
\newcommand{\bea}{\begin{eqnarray}}
\newcommand{\eea}{\end{eqnarray}}

\newcommand{\ed}{\end{document}}

\newcommand{\HH}{{\cal H}}

\newcommand{\QQ}{{\cal Q}}

\begin{document}

\title{$\Z_3$-graded Symmetries in Quantum Mechanics}
\author{Keivan Aghababaei Samani\thanks{E-mail address:
samani@cc.iut.ac.ir}~ and Davood Pooladsaz \\ \\
{\it Department of Physics, Isfahan University of Technology (IUT),}\\
{\it Isfahan 84154, Iran}}
\date{ }
\maketitle

\begin{abstract}
In this paper we consider $\Z_3$-graded topological symmetries
(TSs)~\cite{1,npb} in one  dimensional quantum mechanics. We give
a classification of one dimensional quantum systems possessing
these symmetries and show that different classes correspond to a
positive integer $N$.
\end{abstract}


\section{Introduction}
The notion of $\Z_3$-grading and $\Z_3$-graded structures have
been widely studied in recent years~\cite{int}. In most of these
studies, the $\Z_3$-graded structure is a generalization of a
$\Z_2$-graded structure from a specific point of view, for
example geometrical, group theoretical and algebraical.

In a recent series of papers~\cite{1,npb}we have studied some
$\Z_n$-graded structures called  topological symmetries in quantum
mechanics and explored the algebra of quantum systems possessing
certain topological symmetries. Topological symmetries are
generalizations of supersymmetry from a topological point of view,
i.~e. they share the topological properties of supersymmetry.

A quantum system is said to possess a $\Z_n$-graded (uniform) topological symmetry
(UTS) of type $(m_1,m_2,\cdots,m_n)$ iff the following conditions are satisfied.
    \begin{itemize}
    \item[1.] The quantum system is $\Z_n$-graded. This means that the Hilbert space
    ${\cal H}$ of the quantum system is the direct sum of $n$ of its (nontrivial) subspaces
    ${\cal H}_\ell$, and its Hamiltonian  has a complete    set of eigenvectors with definite
    {\it color} or grading. (A state is said to have a definite color $c_\ell$ iff it
    belongs to ${\cal H}_\ell$);
    \item[2.] The  energy  spectrum is nonnegative;
    \item[3.] Every positive energy eigenvalue $E$ is $m$-fold degenerate, and the
    corresponding eigenspaces are spanned by $m_1$ vectors of color $c_1$, $m_2$ vectors of
    color $c_2$, $\cdots$, and $m_n$ vectors of color $c_n$.
    \end{itemize}
For a system with these properties we can introduce a set of integer-valued
topological invariant, namely
    \be
    \Delta_{ij}:= m_in_j^{(0)} -  m_jn_i^{(0)},
    \label{1.3}
    \ee
where $i,j=1,\cdots,n$ and $n_\ell^{(0)}$ denotes the number of
zero-energy states of color $c_\ell$,~\cite{npb}. Note that the
TS of type $(1,1)$ coincides with supersymmetry and $\Delta_{11}$
yields the Witten index.

Various aspects of topological symmetries are discussed and the
relationship between topological symmetries and
parasupersymmetry, orthosupersymmetry and fractional
supersymmetry is investigated in some recent
articles~\cite{stat,z3,rep}. Also in Ref.~\cite{npb,stat} some
examples of quantum systems possessing topological symmetries are
given, but there is no classification of all quantum systems
possessing topological symmetry until now.

In this article we try to find the most general form of one
dimensional quantum systems with  $\Z_3$-graded topological
symmetries of type $(1,1,1)$ and give a classification scheme for
such systems.

The organization of the paper is as follows. In section 2 we review the algebra of
$\Z_3$-graded topological symmetries of type $(1,1,1)$ and find some general
conditions on the operator in the algebra. In section 3 we find the general solutions
of algebraic relations in one  dimensional quantum systems. Section 4 is devoted to
conclusion and remarks.

\section{$\Z_3$-graded topological symmetries}
The algebra of a $\Z_3$-graded topological symmetry of type $(1,1,1)$ is given by the
following equations
    \bea
    &&[\QQ,H]=0\;,
    \label{a1}\\
    &&\QQ^3=K\;,
    \label{a2}\\
    &&Q_1^3+MQ_1=\frac{1}{2\sqrt 2}(K+K^\dagger)\;,
    \label{a4}\\
    &&Q_2^3+MQ_2=\frac{-i}{2\sqrt 2}(K-K^\dagger)\;,
    \label{a5}\\
    &&[\tau,\QQ]_q=0\;,
    \label{a6}
    \eea
where $H$ is the Hamiltonian of the system, $\QQ$ is the symmetry
generator, $K$ and $M$ are operators which commute with any other
operator in the algebra (furthermore $M$ is hermitian), $\tau$ is
the grading operator with the properties
$\tau^3=1\;,\tau^\dagger=\tau^{-1}$, and  $[.,.]_q$ stands for
$q-commutator$, i.e. $[O_1,O_2]_q=O_1O_2-qO_2O_1$ with
$q=e^{\frac{2\pi i}{3}}$.The operators $Q_1$ and $Q_2$ are related
to $\QQ$ and $\QQ^\dagger$ by the following equations
    \be
    Q_1:=\frac{1}{\sqrt{2}}\:(\QQ+\QQ^\dagger)\;~~~
    {\rm and}~~~Q_2:=\frac{-i}{\sqrt{2}}\:(\QQ-\QQ^\dagger)\;.
    \label{2.1}
    \ee

In view of properties of grading operator of a topological
symmetry the Hilbert space may be expressed as direct sum of
three of its (nontrivial) subspaces $\HH_1$, $\HH_2$ and $\HH_3$.
Each $\HH_i$ is characterized by one of eigenvalues of $\tau$.
Then one can take the following matrix forms for $\QQ$,  $M$  and
$H$.
    \be
    \QQ=\left(\begin{array}{ccc}
             0 &  0  &  D_3\\
             D_1 & 0   & 0 \\
             0  & D_2  & 0
             \end{array}\right)~,~
    M=\frac{1}{2}\left(\begin{array}{ccc}
                       M_1 &0&0\\
                       0&M_2&0\\
                       0&0&M_3
                       \end{array}\right)~,~
                       H=\left(\begin{array}{ccc}
                       H_1 &0&0\\
                       0&H_2&0\\
                       0&0&H_3
                       \end{array}\right)\;,
    \label{rep}
    \ee
where $D_i:\HH_i\rightarrow\HH_{i+1}$ and
$M_i:\HH_i\rightarrow\HH_i$ are some operators which we are going
to determine their explicit forms ($M_i$'s are hermitian).

First of all we note that Eq.~(\ref{a5}) is not an independent
equation. This means that using the matrix forms given by
Eq.~(\ref{rep}) one can see that if Eq.~(\ref{a4}) is satisfied
then Eq.~(\ref{a5}) is fulfilled  trivially. Using the matrix
forms of $\QQ$ and $M$ given by Eq.~(\ref{rep}) and the fact that
$\QQ$ commutes with $H$ and $M$ one gets the following conditions
on $D_i$'s, $H_i$'s and $M_i$'s
    \bea
    &&H_{i+1}D_i=D_i H_i\;,~~i=1,2,3
    \label{6-1}\\
    &&M_{i+1}D_i=D_i M_i\;,~~i=1,2,3
    \label{6}
    \eea
Here and throughout the paper the summation and subtraction  in
subscripts of $D_i$, $M_i$ and $H_i$ is summation and subtraction
modulo 3 respectively; e.~g. we identify $D_4$ with $D_1$ and
$D_0$ with $D_3$. In the same manner Eq.~(\ref{a4}) results in
    \be
    D_i M_i +D_i(D_i^\dagger D_i
    +D_{i-1}D_{i-1}^\dagger)+D_{i+1}^\dagger D_{i+1}
    D_i=0\;,~~i=1,2,3
    \label{7}
    \ee
If we multiply Eq.~(\ref{7}) by $D_i^\dagger$ from the right we
get
    \be
    D_i M_i D_i^\dagger+D_i(D_i^\dagger D_i
    +D_{i-1}D_{i-1}^\dagger)D_i^\dagger+D_{i+1}^\dagger D_{i+1}
    D_iD_i^\dagger=0\;,~~i=1,2,3
    \label{8}
    \ee
The first two terms of Eq.~(\ref{8}) are hermitian so the last
term should  be hermitian as well, i.~e.
    \be
    D_{i+1}^\dagger D_{i+1}D_i D_i^\dagger=D_i D_i^\dagger D_{i+1}^\dagger
    D_{i+1}\;,
    \label{9}
    \ee
or equivalently
    \be
    [D_{i+1}^\dagger D_{i+1},D_i D_i^\dagger]=0\;,~~i=1,2,3
    \label{10}
    \ee

In the same way we can get the following relations from
Eq.~(\ref{6})
    \be
    [M_{i+1},D_i D_i^\dagger]=[M_i,D_i^\dagger D_i]=0\;,~~i=1,2,3
    \label{11}
    \ee

    So there are four mutually commuting
    operators, in $\HH_i$ , $i=1,2,3$ namely $H_i$, $M_i$, $D_{i-1} D_{i-1}^\dagger$,
    $D_i^\dagger D_i$. This fact will help us to find the relation
    of these operators in next section.

\section{One dimensional $\Z_3$-graded TSs}
By {\it ``one dimensional''} we mean that Hilbert spaces $\HH_i$,
$i=1,2,3$ are $L^2(\R)$. We assume that all $D_i$s are nonzero,
because if one of $D_i$s is zero identically then the algebra
given by Eqs.~(\ref{a1})-(\ref{a6}) reduces to the algebra of
$(p=2)$ parasupersymmetry~\cite{rs} and it is well known that a
$(p=2)$ parasypersymmetric quantum system is $\Z_2$-graded not
$\Z_3$-graded~\cite{para}. With the above assumptions we are going
to find the solutions of Eqs.~(\ref{6-1}) -- (\ref{7}). To this
end we use the following lemmas:

\begin{itemize}
\item[]{\bf Lemma 1:} Let $M$ and $N$ be differential operators  of order $m$ and $n$
respectively  acting in the Hilbert space $\HH=L^2(\R)$,
    \bea
    &&M=M_m(x)\frac{d^m}{dx^m}+\cdots+M_1(x)\frac{d}{dx}+M_0(x)\;,
    \label{3.1}\\
    &&N=N_n(x)\frac{d^n}{dx^n}+\cdots+N_1(x)\frac{d}{dx}+N_0(x)\;,
    \label{3.2}
    \eea
where $M_i(x)$ and $N_i(x)$ are complex valued functions of real variable $x$. Then
$[M,N]=0$ iff an operator $D$ exists such that
    \bea
    &&M=a_rD^r+\cdots+a_1D+a_0\;, \label{3.3}\\
    &&N=b_sD^s+\cdots+b_1D+b_0\;, \label{3.4}
    \eea
in which $m=rd$ and $n=sd$ and $d$ is a common divisor of $m$ and
$n$ and  $a_i$'s and $b_i$'s are some complex parameters.
\item[]{\bf Proof:} According to the theory of integrability in Quantum
Mechanics~\cite{int1,int2} we know that for a quantum Mechanical
system described in $N$-dimensional Euclidean space there are  at
most $N$ algebraically independent linear operators commuting
among each other (in such a situation we call the system
integrable). In our system the Hilbert space is $L^2(\R)$ and
this means that the set of linear algebraically independent
operators commuting among each other includes only one element.
Therefore if there are two commuting operators in this system
they are not linearly independent. In other words there must
exist a linear operator $D$ such that Eqs.~(\ref{3.3}) and
(\ref{3.4}) hold.~$\Box$
\item[]{\bf Lemma 2:} Let $M$ and $N$ be differential
operators of order $m$ and $n$ respectively acting in the Hilbert space $\HH=L^2(\R)$,
    \bea
    &&M=M_m(x)\frac{d^m}{dx^m}+\cdots+M_1(x)\frac{d}{dx}+M_0(x)\;,
    \label{3.5}\\
    &&N=N_n(x)\frac{d^n}{dx^n}+\cdots+N_1(x)\frac{d}{dx}+N_0(x)\;,
    \label{3.6}
    \eea
where $M_i(x)$ and $N_i(x)$ are complex valued functions of real variable $x$. if
$MN=0$, then $M=0$ or $N=0$.
\item[]{\bf Proof:} Suppose that $N$ is a non zero differential operator of order $n$
given by Eq.~(\ref{3.6}) (by this assumption we mean that
$N_n(x)$ is a nonzero function of $x$) . By direct calculating of
$MN$ from Eqs.~(\ref{3.5}) and (\ref{3.6}) it is easily seen that
the highest order term is $M_m(x)N_n(x)\frac{d^{m+n}}{dx^{m+n}}$.
As $N_n(x)\ne 0$ and $MN=0$ we conclude that $M_m(x)=0$. By
repeating this procedure we get $M=0$.~$\Box$

\item[]{\bf Lemma 3:} Let $H$ be a second order differential operator of the form
    \be
    H=-\frac{d^2}{dx^2}+V(x)\;,
    \label{3.7}
    \ee
where $V(x)$ is a real valued (nonconstant) function of $x$. Then
there is no first order differential operator $D:=
f(x)\frac{d}{dx}+g(x)$ such that
    \be
    H=a D^2+ b D + c\;,
    \label{3.8}
    \ee
Here $f(x)$ and $g(x)$ are real valued functions of $x$ and $a$, $b$, and $c$ are real
constants.
\item[]{\bf Proof :} Substituting
$D=f(x)\frac{d}{dx}+g(x)$ and $H=-\frac{d^2}{dx^2}+V(x)$ in
Eq.~(\ref{3.8}) we arrive at the following equations
    \bea
    &&af^2(x)+1=0\;,\label{3.82}\\
    &&a\left(f(x)f'(x)+2f(x)g(x)\right)+b f(x)=0\;,\label{3.83}\\
    &&a\left(f(x)g'(x)+g^2(x)\right)+b g(x)+c-V(x)=0\;.\label{3.84}
    \eea
Eqs.~(\ref{3.82}) and (\ref{3.83}) imply that $f(x)$ and $g(x)$
are constants and therefore can not satisfy Eq.~(\ref{3.84})
unless $V(x)$ is constant.~$\Box$
\end{itemize}

Now we are ready to find the solutions of Eqs.~(\ref{6-1}) --
(\ref{7}). In previous section we found out that the operators
$H_i$, $M_i$, $D_{i-1}D_{i-1}^\dagger$ and $D_i^\dagger D_i$ are
mutually commuting in Hilbert space $\HH_i=L^2(\R)$, $i=1,2,3$. At
this stage we put an important restriction on the form of
Hamiltonian. We take the following form for $H_i$s
    \be
    H_i=-\frac{1}{2m_i}\frac{d^2}{dx^2}+V_i(x)\;,i=1,2,3
    \label{3.9}
    \ee
where $V_i(x)$ is real valued well defined potential such that the
spectrum of $H_i$ is non negative. Although by this assumption we
may lose some of mathematical solutions of Eqs.~(\ref{6-1}) --
(\ref{7}), but it seems that we will not lose any physical
solution.

If we confine ourselves to $H_i$ given by Eq.~(\ref{3.9}), then according to Lemma 1
and Lemma 3 $M_i$, $D_{i-1}D_{i-1}^\dagger$ and $D_i^\dagger D_i$ should have the
following forms
    \bea
    D_i^\dagger D_i&=&\sum_{n=0}^N a_n^{(i)} H^n_i\;, \label{3.10}\\
    D_{i-1}D_{i-1}^\dagger &=&\sum_{n=0}^N b_n^{(i)} H^n_i\;, \label{3.11}\\
    M_i&=&\sum_{n=0}^N c_n^{(i)} H^n_i\;, \label{3.12}
    \eea
where $a_n^{(i)}$s, $b_n^{(i)}$s and $c_n^{(i)}$s are real
constants. It should be noted that the above equations does not
mean that $D_i^\dagger D_i$, $D_{i-1}D_{i-1}^\dagger$ and $M_i$
are of the same order because $a_n^{(i)}$s, $b_n^{(i)}$s and
$c_n^{(i)}$s can be zero for some values of $n$. Therefore $N$ is
the greatest $n$ for which at least one of $a_n^{(i)}$,
$b_n^{(i)}$ and $c_n^{(i)}$ is nonzero. In fact as $H_i$'s are
second order differential operators we have $N=Max\{{\cal O}(D_i),
\frac{1}{2}{\cal O}(M_i),i=1,2,3\}$ where ${\cal O}(.)$ stands
for the order of the operator. Substituting Eq.~({\ref{3.12}) in
Eq.~(\ref{6}) and using Eq.~(\ref{6-1}) one gets
$c_n^{(i+1)}=c_n^{(i)}\;,i=1,2,3$. This means that
$c_n^{(1)}=c_n^{(2)}=c_n^{(3)}=:c_n$.

Next  we multiply both sides of Eq.~(\ref{3.10}) by $D_i$ from
the left and use Eq.~(\ref{6-1}) and Lemma 2 and the fact that non
of $D_i$s are zero identically to obtain
    \be
    D_iD_i^\dagger =\sum_{n=0}^N a_n^{(i)} H^n_{i+1}\;.
    \label{3.13}
    \ee
Comparing this equation with Eq.~(\ref{3.11}) one finds that
$b_n^{(i+1)}=a_n^{(i)}\;, i=1,2,3$.

Finally using Eq.~(\ref{3.10}) -- (\ref{3.12}) in Eq.~(\ref{7}) one gets
$c_n=-\sum_{i=1}^3 a_n^{(i)}$. Therefore we can rewrite Eq.~(\ref{3.10}) --
(\ref{3.12}) as
    \bea
    D_i^\dagger D_i&=&\sum_{n=0}^N a_n^{(i)} H^n_i\;, \label{3.14-1}\\
    D_iD_i^\dagger &=&\sum_{n=0}^N a_n^{(i)} H^n_{i+1}\;, \label{3.15-1}\\
    M_i&=&-\sum_{n=0}^N\left(\sum_{j=1}^3 a_n^{(j)}\right)
    H^n_i\;, \label{3.16-1}
    \eea
 These equations show that the
order of  $M_i$, $i=1,2,3$ can not be greater than the greatest
order of $D_i^\dagger D_i$s or $D_iD_i^\dagger$s, $i=1,2,3$ .
Furthermore Eqs.~(\ref{3.14-1}) and (\ref{3.15-1}) imply that
$m_1=m_2=m_3=:m$. This arises from the following fact. Suppose
that we write $D_i$ as $D_i=\sum^N_{i=1}d_i(x)\frac{d^i}{dx^i}$
and calculate $D_i D_i^\dagger$ and $D_i^\dagger D_i$. Obviously
coefficients of the greatest powers of $\frac{d}{dx}$ in $D_i
D_i^\dagger$ and $D_i^\dagger D_i$ are the same. This means that
in the left hand side of Eqs.~(\ref{3.14-1}) and (\ref{3.15-1})
the coefficients of the greatest powers of $\frac{d}{dx}$  are
the same. Therefore this is also true for the right hand side of
these equations. This, in view of Eq.~(\ref{3.9}), means that
$m_1=m_2=m_3=:m$.

From the above calculations we conclude that the  solutions are
classified by the value of $N$. Now we analyze the solutions for
various values of $N$. $N=0$ gives a trivial solution in which
$D_i$s and $M_i$s are constants.

$\bullet$ {\bf Class $N=1$:}

The first nontrivial solution is given by $N=1$. For $N=1$ we have
    \bea
    D_i^\dagger D_i&=&a_0^{(i)}+ a_1^{(i)} H_i\;, \label{3.14}\\
    D_{i-1}D_{i-1}^\dagger &=&a_0^{(i-1)}+ a_1^{(i-1)} H_i\;, \label{3.15}\\
    M_i&=&\left(-\sum_{j=1}^3 a_0^{(j)}\right)-\left(\sum_{j=1}^3 a_1^{(j)}\right)
    H_i\;, \label{3.16}
    \eea
As $H_i$'s are second order differential operators $D_i$'s must be
at most first order differential operators. So we take the
following form for $D_i$'s
    \be
    D_i=f_i(x)\frac{d}{dx}+g_i(x)\;,
    \label{3.17}
    \ee
where $f_i(x)$ and $g_i(x)$ are real valued functions of $x$ (one
can also adopt complex functions of $x$, but without loss of
generality we take $f_i(x)$ and $g_i(x)$ to be real functions of
$x$ reduce the amount of calculations which should be done). Now
substituting $H_i=-\frac{1}{2m}\frac{d^2}{dx^2}+V_i(x)$ and
$D_i=f_i(x)\frac{d}{dx}+g_i(x)$ in Eq.~(\ref{6-1}) we arrive at
the following conditions on $f_i(x)$ and $g_i(x)$.
    \bea
    &&f'_i(x)=0\;,\label{3.19}\\
    &&-\frac{1}{2m}(f''_i(x)+2g'_i(x))+(V_{i+1}(x)-V_i(x))f_i(x)=0\;,
    \label{3.20}\\
    &&-\frac{1}{2m}g''_i(x)+(V_{i+1}(x)-V_i(x))g_i(x)-f_i(x)V'_i(x)=0\;,\label{3.21}
    \eea
Eq.~(\ref{3.19}) implies $f_i=const\;, i=1,2,3$. Thus
Eq.~(\ref{3.20}) takes the following form
    \be
    -\frac{1}{m}g'_i(x)+\left(V_{i+1}(x)-V_i(x)\right)f_i=0\;,
    \label{3.22}\\
    \ee
Using this equation in Eq.~(\ref{3.21}) yields
    \be
    \left(V_{i+1}(x)-V_i(x)\right)g_i(x)=\frac{1}{2}[V'_{i+1}(x)+V'_i(x)]f_i\;,
    \label{3.23}
    \ee
Now we come back to Eqs.~(\ref{3.14}) -- (\ref{3.16}).
Substituting $D_i=f_i(x)\frac{d}{dx}+g_i(x)$ and
$H_i=-\frac{1}{2m}\frac{d^2}{dx^2}+V_i(x)$ in these equations and
using the fact that $f_i(x)=f_i=const.$ one arrives at the
following equations
    \bea
    &&f_i^2=\frac{a_1^{(i)}}{2m}\;,\label{3.25}\\
    &&a_0^{(i)}+a_1^{(i)}V_i(x)=g_i^2(x)-f_i g'_i(x)\;, \label{3.26}\\
    &&a_0^{(i)}+a_1^{(i)}V_{i+1}(x)=g_i^2(x)+f_i g'_i(x)\;, \label{3.27}
    \eea
It is easily verified that in fact  Eqs.~(\ref{3.22}) and
(\ref{3.23}) can be derived from Eqs.~(\ref{3.25}) --
(\ref{3.27}). Therefore in this case the general solution is the
following
   \be
    \QQ=\left(\begin{array}{ccc}
             0 &  0  &  D_3\\
             D_1 & 0   & 0 \\
             0  & D_2  & 0
             \end{array}\right)~~,
    H=\left(\begin{array}{ccc}
                       H_1 &0&0\\
                       0&H_2&0\\
                       0&0&H_3
                       \end{array}\right)\;,
    \label{soll}
    \ee
where $D_i=f_i\frac{d}{dx}+g_i(x)$,
$H_i=-\frac{1}{2m}\frac{d^2}{dx^2}+V_i(x)$ and $f_i$s are real
constants. Also $g_i(x)$, $V_i(x)$, and $f_i$, $i=1,2,3$ should
satisfy Eqs.~(\ref{3.25}) -- (\ref{3.27}). A simple example is
given in the following
   \be
    \QQ=\left(\begin{array}{ccc}
             0 &  0  &  D\\
             g & 0   & 0 \\
             0  & D^\dagger  & 0
             \end{array}\right)~~,
    H=\left(\begin{array}{ccc}
                       H_1 &0&0\\
                       0&H_2&0\\
                       0&0&H_3
                       \end{array}\right)\;,
    \label{sol}
    \ee
where $D=f(\frac{d}{dx}-\alpha x-\beta)$,
$H_1=H_2=\frac{1}{2mf^2}DD^\dagger-a_0$,
 and $H_3=\frac{1}{2mf^2}D^\dagger D -a_0$.The parameters  $\alpha$, $\beta$, $a_0$,
$f$ and $g$ are real constants. Furthermore $M$ and $K$ are given
by the following equations
    \bea
    &&M=-\frac{1}{2}g^2-2mf^2(H+a_0) \;,\label{sol2}\\
    &&K=2mf^2g(H+a_0)\;.\label{sol3}
    \eea
This completes the study of $\Z_3$-graded symmetries
corresponding to $N=1$. For $N>1$, solutions are more
complicated, but with some special choices for the  operators one
can find the solutions easily.

Now we wish to comment on a class of solutions which correspond to
a special choice of operators. In Ref.~\cite{z3} a solution is
given which corresponds to the choice $D_3=(D_2 D_1)^\dagger$.
One can easily verify that this choice satisfies Eqs.~(\ref{6-1})
and (\ref{6}) and $M_2$ is explicitly given by
    \be
    M_2=D_2^\dagger D_2+D_1D_1^\dagger +D_1D_1^\dagger D_2^\dagger
    D_2\;.
    \label{3.29}
    \ee

We use the above choice to find a solution of class $N=2$.

$\bullet$ {\bf Class $N=2$:}

One can easily verify that for $D_3=(D_2 D_1)^\dagger$ the
supercharge $\QQ$ and the Hamiltonian $H$ which satisfy the
algebra of $\Z_3$-graded topological symmetries of type $(1,1,1)$
are given by
   \be
    \QQ=\left(\begin{array}{ccc}
             0 &  0  &  (D_2D_1)^\dagger\\
             D_1 & 0   & 0 \\
             0  & D_2  & 0
             \end{array}\right)~~,
    H=\left(\begin{array}{ccc}
                       H_1 &0&0\\
                       0&H_2&0\\
                       0&0&H_3
                       \end{array}\right)\;,
    \label{soln2}
    \ee
with
    \bea
    &&D_1=f_1 \frac{d}{dx}+g_1(x)\;,
    D_2=f_2\frac{d}{dx}+g_2(x)\label{ex1}\\
    &&H_1=\frac{1}{2mf_1^2}(D_1^\dagger D_1-a_0)=-\frac{1}{2m}\frac{d^2}{dx^2}+V_1(x)\;,
    \label{ex2}\\
    &&H_2=\frac{1}{2mf_1^2}(D_1 D_1^\dagger
    -a_0)=\frac{1}{2mf_2^2}D_2^\dagger D_2=-\frac{1}{2m}\frac{d^2}{dx^2}+V_2(x)\;,\label{ex3}\\
    &&H_3=\frac{1}{2mf_2^2}D_2 D_2^\dagger=-\frac{1}{2m}\frac{d^2}{dx^2}+V_3(x);,\label{ex4}
    \eea
provided that the following conditions are fulfilled
    \bea
    &&g_1^2(x)-f_1g'_1(x)=2mf_1^2V_1(x)+a_0\;,\label{3.30}\\
    &&g_1^2(x)+f_1g'_1(x)=2mf_1^2V_2(x)+a_0\;,\label{3.31}\\
    &&g_2^2(x)-f_2g'_2(x)=2mf_2^2V_2(x)\;,\label{3.32}\\
    &&g_2^2(x)+f_2g'_1(x)=2mf_2^2V_3(x)\;,\label{3.33}
    \eea
Here $f_1$, $f_2$ and $a_0$ are real constants. The operators $M$
and $K$ are given in terms of the Hamiltonian through the
following equations
    \bea
    &&M=-\frac{1}{2}a_0-m(f_1^2+f_2^2(1+a_0))H-2m^2f_1^2f_2^2H^2\;,
    \label{3.34}\\
    &&K=2mf_2^2H(2mf_1^2H+a_0)\;. \label{3.35}
    \eea
It is remarkable that in this class the solution is not unique
because Eqs.~(\ref{3.30}) -- (\ref{3.33}) may have various
solutions for $g_1(x)$ and $g_2(x)$. In fact $V_1(x)$ and
$V_2(x)$ are superpartner potentials and so are $V_2(x)$ and
$V_3(x)$. This result is more or less similar to the result of
some attempts to find an application for $p=2$ parasupersymmetry
in the construction of shape invariant potentials~\cite{bmq}.

\section{Concluding Remarks}
In this paper we  tried to solve the algebraic equations of
$\Z_3$-graded topological symmetries of type $(1,1,1)$. Working
in the framework of one dimensional quantum mechanics and taking a
special form for the Hamiltonian of the system (Eq.~(\ref{3.9}))
we obtained a classification for quantum mechanical systems
possessing these types of symmetries which are defined in
Refs.~\cite{1,npb}. It is shown that the classification is given
by a positive integer, $N$. For $N=1$ and $N=2$  the solutions
obtained explicitly.

One may apply the method of this article to two dimensional
Quantum mechanical systems as well. In this case the Hilbert
spaces $\HH_i$s are $L^2(\R^2)$. This implies that there are two
linear algebraically independent commuting operators in $\HH_i$
which one of them is Hamiltonian. Classification of solutions for
two dimensional Quantum systems is the subject of further
investigation.
\section*{Acknowledgment}
We wish to thank  A. Mostafzadeh  and F. Loran for reading the
first draft of the paper and giving valuable comments. Financial
supports of Isfahan University of Technology is acknowledged.

\end{document}